\documentstyle[12pt]{article}
\newcommand{\mysection}[1]{\setcounter{equation}{0}\section{#1}}

\newcommand{\dz}{\delta_Z}
\newcommand{\xg}{x_{\gamma}}
\newcommand{\yg}{y_{\gamma}}
\newcommand{\xz}{x_Z}
\newcommand{\yz}{y_Z}
\newcommand{\zz}{z_Z}
\newcommand{\sm}{SU(2)_L \times U(1)_Y}
\newcommand{\beq}{\begin{equation}}
\newcommand{\eeq}{\end{equation}}
\newcommand{\bea}{\begin{eqnarray}}
\newcommand{\eea}{\end{eqnarray}}
\newcommand{\eq}[1]{eq.~(\ref{#1})}
\newcommand{\rfn}[1]{(\ref{#1})}

\renewcommand{\titlepage}{\clearpage%
\setcounter{footnote}{0}%
\thispagestyle{empty}\pagestyle{plain}\pagenumbering{arabic}%
\kern1mm
\vskip15mm\normalsize}% End of \titlepage tag
\newcommand{\docnum}[1]{\hbox to \hsize{\hskip123mm\hbox{#1}\hss}}
\renewcommand{\date}[1]{\hbox to \hsize{\hskip123mm\hbox{#1}\hss}}
\renewcommand{\title}[1]{\vskip1em\begin{center}\Large\bf#1\end{center}\vskip2.5em}
\renewcommand{\author}[1]{\vskip0.5em{\bf #1}\vskip0.5em}
\newcommand{\inst}[1]{\vskip0.3em{ #1}\vskip0.5em}
\renewcommand{\abstract}{\begin{center}{\Large \bf
Abstract}\end{center}\quotation}

\begin{document}
\def\thefootnote{\fnsymbol{footnote}}
\begin{titlepage}
\docnum{BI-TP 93/47}
\docnum{PM 93/14}
\docnum{October 1993}
\vspace{1cm}
\title
{The Potential of a New Linear Collider for the Measurement
of the Trilinear Couplings among the Electroweak Vector
Bosons\footnote[2]{\normalsize
Supported by Deutsche Forschungsgemeinschaft and the program
PROCOPE of the DAAD for Franco-German scientific collaboration}}
\begin{center}
\author{{\large M.~Bilenky$^{a)}$\footnote[1]
{\normalsize Alexander von Humboldt Fellow. On leave of absence from
Joint Institute for Nuclear Research, Dubna, Russia.}
J.-L.~Kneur$^{b)}$,~F.~M.~Renard$^{b)}$,~D.~Schildknecht$^{a)}$}}
\inst{a)~{\it Department of Theoretical Physics,
University of Bielefeld, 33501  Bielefeld, Germany}}
\inst{b)~{\it Physique Math\'ematique et Theorique, CNRS-URA 768,
Universite Montpellier II, F-34095,
Montpellier Cedex 5, France}}
\end{center}
\vspace{0.5cm}
\begin{abstract}
We study the accuracy to be obtained in measuring trilinear
 $Z^0W^+W^-$ and $\gamma W^+W^-$ couplings in the reaction
 $e^+e^- \to W^+W^-$ at "New Linear Collider"
 energies of $500GeV$ to $1000GeV$.
 We derive simple scaling laws for the sensitivity in the
 measurement of these couplings. For most couplings the
 sensitivity increases as $\sqrt{L\cdot s}$, where $L$
 denotes the integrated luminosity and $\sqrt s$ denotes the
 $e^+e^-$ center-of-mass energy. Detailed
 investigations based on various
 fits confirm these scaling laws and show
 that an accuracy of the order of the standard radiative
 corrections can be reached at
 the NLC for the design values of the luminosity.
\end{abstract}
\end{titlepage}
\setcounter{page}{1}
\def\thefootnote{\arabic{footnote}}
\setcounter{footnote}{0}
%          MAIN TEXT OF THE PAPER
\mysection{Introduction}
\label{intro}
Our present empirical knowledge on electroweak phenomena is largely
confined to vector-boson-fermion interactions. Our knowledge
on other properties of the weak vector bosons, apart from their
masses, is rather limited.
{\it Indirectly}, the electroweak precision data\cite{ewdata}
imply restrictions on non-standard bosonic self-interactions via
model-independent bounds
(see e.g. \cite{minda})
on radiative corrections.
{\it Direct} investigations of these couplings
at future colliders will be indispensable, however,
for a full understanding of the electroweak
interactions.
\par
We will discuss
the measurements
of the {\it trilinear couplings} among
vector bosons in the process
$e^+e^- \to W^+W^-$~~\footnote{We refer to, e.g.,
\cite{quatr} for a discussion of the quadrilinear couplings.
Study of the self interactions of the vector bosons in other processes
at future hadron colliders and in $e\gamma$ and
$\gamma\gamma$ interactions
can be found in \cite{ppano} and \cite{geano}, respectively.}.
The first experiments on this reaction
will be carried out at LEP2 in the near future.
The potential of
LEP2 for measuring the trilinear couplings among the vector bosons
was analysed recently \cite{BKRS93} in some detail
(see also \cite{lep2w}
for earlier studies). At an $e^+e^-$ energy
of about $190 GeV$ and with an integrated luminosity of $500 pb^{-1}$
an accuracy of order 0.1 for the determination of the trilinear
couplings can be reached.
This will improve present direct constraints on the $W^+W^-\gamma$
vertex from $p\bar p$ colliders
\cite{UA2} by more than one order of magnitude. Indirect bounds
on non-standard couplings estimated from loop corrections
\cite{DGHM92},
\cite{HISZ93}
to electroweak precision data will be improved by
factors ranging from about
2 to an order of magnitude.
{}From these results, one
will be able to rule out (or find) drastic deviations from
standard-model predictions. Measurements at LEP2 will not be
sufficient, however, as the appropriate scale for the precision
to be aimed at
is determined by the deviations from
tree-level standard trilinear couplings
induced by (standard) radiative corrections. This scale is of the
order $10^{-2}$ to
$10^{-3}$ \cite{RC92}. The question naturally
arises, whether this level of sensitivity can be reached in the energy
range and with the luminosities
now envisaged for an $e^+e^-$ collider.
\par
In the present work, our recent analysis for the LEP2 energy range
\cite{BKRS93} is extended to the energies of $500 GeV$ and $1000 GeV$
of a future linear $e^+e^-$-collider (NLC) \cite{NLC1,NLC2}.

\mysection{Theoretical restrictions on non-standard couplings}
\label{theorconstr}
\par
For the present analysis, we disregard the possibility of CP-violating
couplings\footnote{A unique procedure
to search for CP-violating $Z^0W^+W^-$
couplings via comparison of appropriate $W^-$ and $W^+$
spin-density-matrix elements was presented in \cite{GSR91}.}.
Assuming
$C-$ and $P-$invariant photon interactions, we can
effectively describe
\cite{efflagr}
the $\gamma W^+W^-$ and $Z^0W^+W^-$ couplings by
the Lagrangian
\cite{BKRS93}
\bea
&&L = -ie[A_\mu
(W^{-\mu \nu}W_\nu^+ -W^{+\mu\nu}W^-_\nu)+ F^{\mu\nu}W_{\mu}^+W_{\nu}^-]
- iex_\gamma F^{\mu\nu}W_{\mu}^+W_{\nu}^- \nonumber\\
&&-ie({\rm ctg}\theta_W + \dz)
[Z_\mu(W^{-\mu \nu}W_\nu^+ -W^{+\mu\nu}W^-_\nu)+ Z^{\mu\nu}W_{\mu}^+W_{\nu}^-]
- ie\xz Z^{\mu\nu} W_{\mu}^+W_{\nu}^- \nonumber\\
&&+ie{{y_\gamma}\over{M^2_W}}F^{\nu\lambda}W^-_{\lambda\mu}W^{+\mu}_\nu
+ie{{\yz}\over{M^2_W}}Z^{\nu\lambda}W^-_{\lambda\mu}W^{+\mu}_\nu \nonumber\\
&&+{{e\zz}\over {M^2_W}}
\partial^\alpha \hat Z_{\rho}^{\sigma}
(\partial^\rho W_{\sigma}^-W_{\alpha}^+
-\partial^\rho W_{\alpha}^-W_{\sigma}^+
+\partial^\rho W_{\sigma}^+W_{\alpha}^-
-\partial^\rho W_{\alpha}^+W_{\sigma}^-),
\label{lagr}
\eea
where $F_{\mu\nu},~Z_{\mu\nu},~W^{\pm}_{\nu}$ are Abelian field-strength
tensors for photon, $Z$ and $W^{\pm}$ bosons, respectively, and
\beq
 {\hat Z}_{\mu\nu} = {1 \over 2}
\epsilon_{\mu\nu\sigma\rho}Z^{\sigma\rho}.\nonumber
\label{tensor}
\eeq
The parameter $\dz$ describes a deviation of the $Z^0W^+W^-$
overall coupling from its standard value. Non-zero values of $\xg$
and $\xz$ parametrize potential deviations in the electromagnetic
and weak dipole couplings from the standard model predictions, and
$\yg,~\yz$ denote the strengths of non-standard dimension-six
quadrupole interactions of the $W^\pm$. The coupling $z_Z$ describes
a $CP$-conserving, but $C$- and $P$-violating, so-called anapole
coupling of the $Z^0$ to the $W^\pm$.
\par
The Lagrangian \rfn{lagr} contains the trilinear interactions
of the standard model {\it at tree level}
%\cite{SM}
for the special case of
\beq
\dz = \xg = \xz = \yg = \yz = \zz =0.
\label{smcoup}
\eeq
The symmetry and renormalizability requirements formulated in the
standard electroweak theory lead to the restriction \rfn{smcoup}.
In the most general
phenomenological model-independent analysis
of experimental data
(assuming $CP$ invariance)
all six parameters in \eq{lagr} must
be treated as independent ones.
There are nevertheless theoretical
as well as practical reasons to reduce the number of free
trilinear (and quadrilinear) couplings
by additional constraints
based on $SU(2)$-symmetry requirements
(refs. \cite{MSS86}-\cite{SCH86}),
thus excluding the theoretically most disfavoured deviations
from the standard electroweak theory.
%\cite{MSS86,KRS87,,NSS87BKS88,SCH86},
\par
The restrictions from $SU(2)$ symmetry on the trilinear
couplings in \rfn{lagr} can be simply reproduced
\cite{BKRS93} by performing a transformation from the $\gamma Z^0$
to the $W^3B$ (or, alternatively, the $\gamma W^3$ current-mixing)
base in the Lagrangian \rfn{lagr}.
Here, we briefly summarize the results and refer to
refs. \cite{BKRS93} and \cite{MSS86} to \cite{SCH86} for details.
\par
Excluding intrinsic $SU(2)$ violation, i.e., requiring restoration of
$SU(2)$ symmetry in the decoupling limit of the hypercharge $(B_\mu)$
field, $e = s_W = 0$, implies the condition \cite{MSS86}
\beq
x_Z= - {{s_W}\over {c_W}} x_\gamma
\label{xgxz}
\eeq
(where $s_W=e/g_W$ denotes the sine of the weak mixing angle and
$c^2_W \equiv 1- s^2_W$), while allowing for
\beq
x_\gamma~~,~~\delta_Z~~,~~y_\gamma~~,~~y_Z \not= 0.
\eeq
The number of free non-standard couplings is further
reduced, if $SU(2)$ symmetry is imposed
on the quadrupole interaction, implying
\cite{KRS87}
%\cite{MSS86,KRS87,NSS87,BKS88,SCH86},
\beq
\yz= {{c_W}\over {s_W}} \yg .
\label{ygyz}
\eeq
Further requirements, such as a fairly decent high-energy behaviour
of the tree-level amplitudes for
the scattering of vector bosons on each other, lead to additional
constraints \cite{NSS87,BKS88}.
The various constraints are collected in Table 1 taken
from ref. \cite{BKRS93}.
%----------------------------------------------------------
% begin of the table 1
\begin{table}[thbp]
\begin{center}
\begin{tabular}{||l|c|l||}
\hline
 & number & \\
Symmetry & of & Couplings and constraints \\
 & param. & \\
\hline\hline\hline
Lorentz-invariance & 3 & $\dz,~\xg,~\xz$ \\
C-,~P-invariance   &   &                 \\
\hline\hline
Exclusion of intrinsic SU(2) violation; & 2 &
$\dz,~\xg;$ \\
local $\sm$, by including & & $\xz = - {{s_W} \over {c_W}} \xg$   \\
dim.-6 Higgs interaction via $L_{W\phi}, L_{B\phi}$ & &  \\
\hline\hline
Exclusion of $s^2$ terms in $W^+W^-$, etc.& 1
& $\xg;$ \\
scattering; & & $\dz= {{\xg} \over {s_W c_W}},~
\xz = - {{s_W} \over {c_W}} \xg$  \\
Exclusion of $\Lambda^4$-divergence in $\rho$; & & \\
$L_{W\phi}$ only & & \\
\hline\hline
$L_{B\phi}$ only & 1 & $\xg;$ \\
 & & $\dz = 0,~\xz = -{{s_W}\over {c_W}} x_\gamma$ \\
\hline\hline
$L_{W \phi} + L_{B \phi}$ & 1 & $\xg;$ \\
 & &$\dz = {{\xg}\over {2 s_W c_W}}$,~$\xz = - {{s_W} \over {c_W}} \xg$  \\
\hline\hline
$L_{W \phi} - L_{B \phi}$ or, alternatively,& 1 & $\dz;$\\
$SU(2)_W \times SU(2)_V \times U(1)_Y$ & & $\xg = \xz = 0$ \\
\hline\hline\hline
Lorentz-invariance, & 5 & $\dz, \xg, \xz, \yg, \yz$  \\
 C-, P-invariance & &  \\
\hline\hline
 Exclusion of & 4 & $ \dz,~\xg,~\yg,~\yz;$ \\
intrinsic $SU(2)$ violation & & $\xz = - {s_W \over c_W} \xg$  \\
\hline\hline
Local $\sm$; & 3 & $\dz,~\xg,~\yg$; \\
$L_{W \phi}, L_{B \phi}$ and quadrupole, $L_W$ &   &
$\xz = - {{s_W} \over {c_W}} \xg,~\yz = {{c_W} \over {s_W}} \yg$ \\
\hline\hline
Local $\sm$ & 2 & $\xg,~\yg;$ \\
$L_{W \phi}$ and $L_W$ & & $\dz = {{\xg}\over {s_W c_W}},
{}~\xz= - {{s_W} \over {c_W}} \xg,~\yz =  {{c_W} \over {s_W}} \yg$ \\
\hline\hline
Local $\sm$ & 1 & $\yg;$ \\
$L_W$ only & & $\yz = {{c_W} \over {s_W}} \yg,~\dz = \xg = \xz = 0$ \\
\hline\hline\hline
Lorentz-invariance, & 6 & $\dz,~\xg,~\xz,~\yg,~\yz,~\zz$  \\
C-,~P-violation & & \\
\hline
\end{tabular}
\caption[{\bf Table 1}]
{\it Constraints on the $\gamma W^+W^-$ and $Z^0 W^+ W^-$ couplings
in Lagrangian \eq{lagr}.
The second column
shows the number of free parameters. The free parameters and
the constraints defining the remaining parameters are displayed
in the third column.
In the upper part of the Table, dimension-six quadrupole
terms are excluded by assumption $(\yg = \yz = 0)$,
while in the lower part of the Table such terms are allowed.}
{\label{tab1}}
\end{center}
\end{table}
%-----------------------------------------------------------------------------
% end of the table 1
\par
The three-free-parameter $(\dz, \xg, \yg )$
interaction Lagrangian with the constraints \rfn{xgxz} and \rfn{ygyz} on
$x_Z$ and $y_Z$ may be incorporated \cite{BKRS93,KKS93}
into a Lagrangian which is
invariant under local $SU(2)$ transformations\footnote {A complete
list of such $\sm$- operators can be found e.g. in ref. \cite{BW86}}.
While this
embedding of the interactions into such a framework is irrelevant
for the (tree-level) phenomenology of the reaction
$e^+ e^- \rightarrow W^+W^-$, it is
of importance insofar as it provides an example of how
non-standard couplings can coexist with LEP1 precision data:
the linearly realized local $SU(2)$ symmetry
assures \cite{DGHM92,HISZ93} an at most logarithmic
dependence on the cut-off in one-loop corrections to LEP1
observables and decoupling of ``new physics'' effects.
In the case of the dimension-six quadrupole interactions, local $SU(2)$
symmetry is simply obtained \cite{KRS87} by appropriate use of the
non-Abelian field tensor for the $W$
field. For the
dimension-four trilinear interactions, however, the introduction
of non-standard Higgs interactions
\cite{DGHM92,HISZ93,GR93,KKS93}
is essential.
The basic Lagrangian takes the form \cite{BKRS93,KKS93}
\bea
L =
{2\over M^2_W} (\xg - \dz s_W c_W ) L_{B\phi} + {2\over M_W^2}
\dz s_W c_W L_{W\phi} \nonumber \\
+ e{y_\gamma \over{s_W M^2_W}} L_W,
\label{su2lagr}
\eea
with
\bea
&&L_{B\phi} =
i {e \over {2c_W}}B^{\mu\nu}(D_{\mu}\phi)^{\dagger}(D_{\nu}\phi),\nonumber \\
&&L_{W\phi} =
i{e \over {2s_W}}\vec{{\rm w}}^{\mu\nu} (D_{\mu}\phi)^{\dagger}
\vec{\tau}\cdot (D_{\nu}\phi), \\
&&L_W = {1\over 6} \vec{{\rm w}}^{\mu\lambda} (\vec{{\rm w}}
_{\lambda\nu} \times\vec{{\rm w}}^\nu_{~\mu} ), \nonumber
\label{operators}
\eea
where $D_\mu$ denotes the covariant derivative
\beq
D_\mu =\partial_\mu + i
{e\over s_W} {{{\buildrel\rightharpoonup\over\tau}}\over 2}
{\buildrel\rightharpoonup\over W}_\mu + i{e\over c_W}B_\mu Y
\label{deriv}
\eeq
and $\vec{{\rm w}}_{\mu\nu}$ the non-Abelian field tensor
\beq
\vec{{\rm w}}_{\mu\nu} =
\partial_{\mu}\vec{W}_{\nu}
-\partial_{\nu}\vec{W}_{\mu} -{e\over s_W} \vec{W}_{\mu} \times
\vec{W}_{\nu}.
\label{tensor1}
\eeq
Upon passing to the physical $\gamma$ and $Z^\circ$ fields in
\eq{su2lagr}, for the trilinear couplings, one
recovers Lagrangian \rfn{lagr} with
$(\dz , \xg , \yg )$ as free parameters and the
constraints \rfn{xgxz} and \rfn{ygyz} for $x_Z$ and $y_Z$, respectively.
\par
Various specific cases of the Lagrangians \rfn{lagr}, \rfn{su2lagr},
corresponding to different constraints among the couplings,
are collected in Table 1 taken from \cite{BKRS93}.

\mysection{Scaling laws for the bounds on
non-standard couplings}
The helicity amplitudes for the process $e^+e^- \rightarrow W^+W^-$
corresponding to the general Lagrangian \rfn{lagr}
were given in Table 3 of ref.\cite{BKRS93}. Here we
restrict ourselves to a brief discussion of the high-energy
dependence
of the cross section for the production of $W^\pm$ bosons
with correlated helicities. We will see that the sensitivity
for the determination of non-standard couplings in the
high-energy limit can be represented by a simple formula
in terms of the $e^+e^-$ energy and the integrated
$e^+e^-$ luminosity.
\par
We consider the cross section, $\sigma^{AA^\prime} (s)$,
where $(A,A^\prime)= (L,T)$,
for the production of $W^+W^-$-pairs of definite helicities.
The indices $T$ and $L$
refer to longitudinal ($W^\pm$-helicity $0$) and transverse
($W^\pm$-helicity $\pm 1$) polarizations, respectively.
We assume that the non-standard couplings,
$\dz, \xg, \xz,$ etc., are sufficiently small to be treated
in the  linear approximation, i.e., all purely non-standard
contributions (proportional to $\delta^2_Z, x^2_\gamma, x^2_Z,
\ldots$)
are neglected.
The cross section $\sigma^{AA^\prime} (s)$ then becomes
\beq
\sigma^{AA^\prime} (s) =
\sigma^{AA^\prime}_0 (s)
+ \sum^6_{i = 1} x_i \Delta_i^{AA^\prime} (s),
+ O(x_i^2),
\label{xsect}
\eeq
where $\sigma^{AA^\prime}_0 (s)$ denotes the standard-model
cross section, and the parameters
$x_i$
stand for the six
non-standard couplings in \eq{lagr},
$(x_1,\ldots , x_6)\equiv (\delta_Z, x_\gamma ,\ldots , z_Z)$.
The asymptotic $(s \gg 4 M^2_W)$ energy dependence of
the coefficients $\Delta_i^{AA^\prime} (s)$ in \eq{xsect}
is easily obtained from the expressions for the helicity amplitudes
explicitly represented in Tables 3 and 4 of ref. \cite{BKRS93}.
The result is displayed
in Table 2\footnote{We thank Dr. A. Pankov for a useful
discussion related to the content of Table 2.}.
Due to (energy-independent) selection rules for the
dipole, quadrupole and anapole couplings, certain coefficients
are vanishing in \eq{xsect} and, consequently, certain entries
in Table 2 are absent.
%-----------------------------------------------------------------------------
% begin of the table 2
\begin{table}[thbp]
\begin{center}
\begin{tabular}{||c|c|c|c|c||}
\hline
   & TT           &           TT & LL & LT \\
   & $\displaystyle \tau = \tau'$ & $\displaystyle \tau =-\tau'$ &    &    \\
\hline\hline
  & & & & \\
$\sigma_0^{AA^\prime}(s)$ & $\displaystyle s^{-3}$ & $\displaystyle s^{-1}$
& $\displaystyle s^{-1}$ & $\displaystyle s^{-2}$  \\
  & & & & \\
\hline
  & & & & \\
$\Delta_{\dz}^{AA^\prime}(s)$       & $\displaystyle s^{-2}$ & & $const.$
& $\displaystyle s^{-1}$    \\
  & & & & \\
\hline
  & & & & \\
$ \Delta_{\xg,x_Z}^{AA^\prime}(s)$  &    &         & $const.$
& $\displaystyle s^{-1}$    \\
  & & & & \\
\hline
  & & & & \\
$ \Delta_{y_\gamma,~y_Z}^{AA^\prime}(s)$  &$\displaystyle s^{-1}$
&    &
& $\displaystyle s^{-1}$    \\
  & & & & \\
\hline
  & & & & \\
$ \Delta_{z_Z}^{AA^\prime}(s)$        &    &      &      & const          \\
  & & & & \\
\hline
\end{tabular}
\caption[{\bf Table 2}]
{\it The asymptotic ($s \gg 4 M_W^2$) energy dependence
of the standard model cross section (first row) and
of the various non-standard contributions
$\Delta_i^{AA^\prime}(s)$ in \eq{xsect}.
If both vector bosons, $W^+$ and $W^-$, have transverse polarization
$(TT)$, the two cases of equal
$(\tau =\tau^\prime = \pm 1)$ and opposite
$(\tau =- \tau^\prime = \pm 1)$ helicities have different high-energy
behaviour, as indicated.}
{\label{tab2}}
\end{center}
\end{table}
%-----------------------------------------------------------------------------
% end of the table 2
\par
Using Table 2, the energy dependence of the sensitivity for the
measurement of the non-standard couplings is easily derived.
Restricting ourselves to small non-standard contributions to
$\sigma^{AA^\prime}(s)$, an assumption already introduced in \eq{xsect},
the statistical error of a measurement of the cross-section
$\sigma^{AA^\prime}(s)$
may be approximated by the error corresponding to the number of events
calculated from the standard cross section,
$\sigma_0^{AA^\prime}(s)$,
via
\beq
\delta N^{AA^\prime}=const.\sqrt{L\cdot \sigma_0^{AA^\prime}(s)}.
\label{3.2}
\eeq
In \rfn{3.2}, $L$ denotes the integrated $e^+e^-$ luminosity.
An energy-independent proportionality constant, substantially
larger than unity\footnote{Actually, the proportionality constant
depends on $A, A^\prime$. Its absolute value is irrelevant,
however, for the relative values of the sensitivity (as a
function of energy and luminosity) under consideration in the
present section.}, appears in \rfn{3.2}, since in
actual experiments the helicity information can only be extracted
by an analysis of the $W^\pm$ decay distributions (in their
respective rest frames, compare ref. \cite{BKRS93} and section 4
of the present paper). Equating the statistical error \eq{3.2}
with the number of non-standard excess events predicted by \eq{xsect},
\beq
\sqrt{L \sigma_0^{AA^\prime}(s)}=
const. L \cdot \mid \sum_i x_i\Delta_i^{AA^\prime} (s)\mid,
\label{3.3}
\eeq
allows one to determine (the energy dependence of) the sensitivity of
a measurement of the couplings $x_i$. Asymptotically, for given
helicities, $A, A^\prime$, according to Table 2, those non-standard
couplings are dominant in \rfn{3.3} which are associated with
either a constant (in case of $\delta_Z, x_\gamma, x_Z$ and $z_Z$)
or else an $s^{-1}$ (in case of $\yg,~y_Z$)
energy dependence. Accordingly, from \eq{3.3} the sensitivity
for the measurement of $x_i$ (defined by the inverse of the
magnitude of $x_i$) is determined by the proportionality\footnote
{When deriving \rfn{3.4} from \rfn{3.3}, in general a sum of several
$x_i$ with energy-independent coefficients will appear in the numerator
of \rfn{3.4}. The proportionality \rfn{3.4} then follows immediately.}
\beq
{1\over{\mid x_i\mid}} \sim \Delta_i^{AA^\prime} (s)
\sqrt{{L\over{\sigma_o^{AA^\prime}(s)}}},
\label{3.4}
\eeq
where for a definite choice of $x_i$, the right-hand side has to be
evaluated for those helicities $A, A^\prime$, for which, according
to Table 2, the cross section $\sigma^{AA^\prime}(s)$ is dominated
by the contribution proportional to $x_i$, i.e., we have to use
the following correspondance when evaluating \eq{3.4}:
\bea
(y_\gamma, y_Z) & \to & TT (\tau = \tau^\prime),\nonumber\\
(\delta_\gamma, x_\gamma, x_Z) & \to & LL, \label{3.5}\\
z_Z & \to & LT.\nonumber
\eea
\par
Explicitly, from eqs. \rfn{3.4} and \rfn{3.5}, using Table 2, we
find the results
displayed in Table 3.\break
%------------------------------------------------------------------
% begin of the table 3
\begin{table}[thbp]
\begin{center}
\begin{tabular}{||c|c||}
\hline
 coupling     & sensitivity    \\
\hline\hline
  & \\
$\dz,~\xg,~x_Z,~\yg,~y_Z$ & $\sqrt{s \cdot L}$ \\
  & \\
\hline
  & \\
$z_Z$ & $s\sqrt{L}$ \\
  & \\
\hline
\end{tabular}
\caption[{\bf Table 3}]
{\it Sensitivity for the determination of the non-standard
couplings as a function of the integrated $e^+e^-$ luminosity, $L$, and
the square of the center-of-mass energy, $s$.}
{\label{tab3}}
\end{center}
\end{table}
%-----------------------------------------------------------------------------
% end of the table 3
According to this Table, the sensitivity
for the anapole interaction increases as $s\sqrt L$, while for
all other couplings it increases as $\sqrt{s\cdot L}$.
We draw attention to the fact that the simple behaviour
\rfn{3.4}, \rfn{3.5} leading to Table 3, according to \rfn{xsect},
is based on the linear approximation in $x_i$. The neglect of terms of
order $x_i^2\cdot (s/M^2_W)$ will break down at sufficiently large values
of s, even for small values of $x_i$. We will see, however, that
the sensitivity is well described by Table 3 in the range of energies
and luminosities to be considered explicitly below. Finally, scaling
laws can be different if the full $W^\pm$ helicity information is not
taken into account, as helicity information explicitly enters our error
analysis (e.g., in \rfn{3.2}).
\par
In summary, once the sensitivity is known for a specific energy
(sufficiently above threshold, $s >> 4 M^2_W$), and luminosity,
the dependence from Table 3 may be used to predict the sensitivity
for any other (asymptotic) energy. Explicit
%------------------------------------------------------------------
% begin of the table 4
\begin{table}[thbp]
\begin{center}
\begin{tabular}{||c|c|c|c||}
\hline
   &  L~$[pb^{-1}]$ & $\displaystyle \dz,\xg,x_Z,\yg,y_Z$&
 $\displaystyle z_Z$ \\
\hline\hline
 & & & \\
 LEP 200 & 500   & ~   & ~    \\
 ~       & ~               & $\sim$ 11 & $\sim$ 28  \\
 NLC 500 & 10000 & ~   & ~    \\
 ~       & ~               & $\sim$~ 4 & $\sim$ 8  \\
 NLC 1000& 44000 & ~   & ~    \\
 & & & \\
\hline
\end{tabular}
\caption[{\bf Table 4}]{\it Increase in sensitivity according to \rfn{3.4},
\rfn{3.5} and Table 3..
The assumed integrated luminosities are listed in the second column.}
{\label{tab4}}
\end{center}
\end{table}
%-----------------------------------------------------------------------------
% end of the table 4
%-----------------------------------------------------------------
numerical results for the increase in sensitivity from LEP2 to the
NLC\footnote{The integrated luminosity of $10 fb^{-1}$ at $500 GeV$
corresponds to $\sim 10^7$ sec. of operation for the Palmer F design of
a linear collider
\cite{NLC1}. This option of a linear collider has a narrow
energy distribution around $500 GeV$.}, obtained by evaluating
the formulae of Table 3, are presented in Table 4. In section 4, these
results will be compared with the results of numerical
simulations performed
for the cases listed in Table 4. As expected from the above
derivations, we will find perfect agreement with the simple
scaling laws of Tables 3 and 4.
\mysection{Simulation of the experimental determination of the
trilinear couplings.}
In the numerical analysis of the precision to be
expected for the measurements of the trilinear couplings, we
only consider events of the type
\beq
e^+e^- \to W^+W^- \to \bigg\{
\begin{array}{ccc} e^\pm \nu_e & + & 2~{\rm jets}, \\
         \mu^\pm \nu_\mu & + & 2~{\rm jets}.
\end{array}
\label{4.1}
\eeq
For these events identification of the charge of the $W$-bosons
will be simple. We will not consider here the possibility of
electron (positron) beam polarization (see e.g. \cite{BAR92}).
With the luminosities of Table 4, when
assuming standard tree-level amplitudes for (on-shell) $W^\pm$
production in the angular range of
\beq
- 0.98 \leq \cos \vartheta \leq 0.98,
\label{4.2}
\eeq
one obtains
\beq
\begin{array}{crcccr}
\sim & ~2900 & {\rm events} & (E_{e^+e} & = & 190 GeV), \\
\sim & 13200 & {\rm events} & (E_{e^+e} & = & 500 GeV), \\
\sim & 14000 & {\rm events} & (E_{e^+e} & = & 1000 GeV).
\end{array}
\label{4.3}
\eeq
\par
Following ref.\cite{BKRS93},
we simulate the two-step procedure for the analysis of the data
suggested therein.
\par
According to this procedure, in a first step, all observables,
i.e., the differential cross section as well as the various
single-particle spin-density-matrix elements and the $W^+ W^-$ spin
correlations, are to be determined.
Only the well known $V-A$ charged-current
weak interaction enters, when extracting the spin properties
of the produced $W^\pm$ bosons from the measured $W^\pm$ decay
distributions. Consequently,
as far as the $W^\pm$ production
process is concerned,
this first step in the data analysis
is entirely {\it model independent}.
Once actual data will become available, a comparison with the
results of the model-independent analysis with (standard and/or
non-standard) theoretical predictions is to be carried out.
In the subsequent second step the trilinear couplings are
to be determined by a
fitting procedure, using the differential cross section and the
spin-density matrix elements as the empirical input.
\par
In our simulation of the first step of the data analysis we
generate "data" for the three-fold differential cross sections
\beq
{{d\sigma}\over{d\cos\vartheta d\cos\theta_1 d\phi_1}}\quad ,\quad
{{d\sigma}\over{d\cos\vartheta d\cos\theta_2 d\phi_2}}\quad ,\quad
{{d\sigma}\over{d\cos\vartheta d\cos\theta_1 d\cos\theta_2}}
\label{dxsect}
\eeq
in accordance with the standard model, assuming the integrated
luminosities given in Table 4 and the corresponding event numbers
\rfn{4.3}.
In \eq{dxsect}, $\vartheta$ denotes the $W$-production
angle and $\theta_i,\phi_i (i=1,2)$ denote the polar and azimuthal
$W^\pm$ decay angles. As a result of the fit, we obtained
"data" in the form of standard-model values with statistical
errors
\footnote{Statistical errors only are taken into
account. A discussion of systematic errors is beyond the scope
of the present investigation. Some studies of
systematical uncertainties can be found in \cite{exp}.}
for all above-mentioned observables.
\par
In the second step,
the differential cross sections and the density-matrix elements
(with their errors)
serve as the input for the determination of the trilinear couplings.
Fits were carried out for the different parametrizations presented
in Table 1.
The conclusion of ref.[6] that helicity information
is particularly important in multi-parameter fits and that it frequently
improves bounds by factors of the order of 2, or by even larger factors,
was found to remain valid at NLC energies.
\par
The resulting bounds on the non-standard couplings
are collected in Table 5,
\small
%-----------------------------------------------------------------------------
% begin of the table 5
\begin{table}[thbp]
\begin{center}
\begin{tabular}{||c|c|c|c|c|c|c||}
\hline
   & $\delta_Z$ & $x_\gamma$ & $x_Z$ & $y_\gamma$ & $y_Z$ & $z_Z$ \\
\hline\hline\hline
% 31
 & $-1.19 \div 1.27$ & $-0.29 \div 0.95$ & $-2.48 \div 2.13$ &  &  &  \\
\cline{2-4}
3 &$-0.15 \div 0.11$ & $-0.01 \div 0.08$ & $-0.17 \div 0.23$ & 0 & 0 & 0 \\
\cline{2-4}
   &$-0.052 \div 0.041$ &$-0.003 \div 0.023$ &$-0.06 \div 0.08$ & & & \\
\hline\hline
% 21
  &$-0.32~ \div 0.20~$ & $-0.27~ \div 0.87~$ & & & & \\
\cline{2-3}
 2 &$-0.042 \div 0.024$ & $-0.010 \div 0.076$ &
$\displaystyle - {s_W \over c_W} x_\gamma$&
0 & 0 & 0 \\
\cline{2-3}
   &$-0.011 \div 0.010$ & $-0.002 \div 0.021$ & & & & \\
\hline\hline
% 11
   & & $-0.04~ \div 0.04~$ & & & &  \\
\cline{3-3}
 1 &
$\displaystyle {x_\gamma \over{s_W c_W}}$ & $-0.003 \div 0.003$
 &
$\displaystyle -{s_W\over c_W} x_\gamma$ & 0 & 0 & 0 \\
\cline{3-3}
   & & $-0.001 \div 0.001$ & & & & \\
\hline\hline
% 12
  & & $-0.17~ \div 0.21~$ & & & & \\
\cline{3-3}
 1 & 0 & $-0.006 \div 0.007$ &
$\displaystyle - {s_W\over c_W} x_\gamma$ & 0 & 0 & 0 \\
\cline{3-3}
   & & $-0.002 \div 0.002$ & & & & \\
\hline\hline
% 13
  & & $-0.06~ \div 0.07~$ & & & & \\
\cline{3-3}
 1 &
$\displaystyle {x_\gamma \over{2s_W c_W}}$ &$-0.004 \div 0.005 $
&
$\displaystyle -{s_W \over c_W} x_\gamma$ & 0
& 0 & 0 \\
\cline{3-3}
   & & $-0.001 \div 0.001$ & & & & \\
\hline\hline
% 14
%  & & $-0.12~ \div 0.13~$ & & & & \\
%\cline{3-3}
% 1 & 0 &$-0.005 \div 0.006$ & 0 & 0 & 0 & 0 \\
%\cline{3-3}
%   & &$-0.001 \div 0.001$ & & & & \\
%\hline\hline
% 15
  & $-0.11~ \div 0.12~$ & &  &  &  &  \\
\cline{2-2}
 1 & $-0.012 \div 0.014$  & 0 & 0 & 0 & 0 & 0 \\
\cline{2-2}
   &$ -0.004 \div 0.005$ & & & & & \\
\hline\hline\hline
% 51
  & $-4.78 \div 4.55$ & $-0.45~ \div 1.09~$ & $-6.87 \div 7.08$ &
$-0.56 \div 0.77$ & $-2.11 \div 2.19$ &  \\
\cline{2-6}
 5 &$-0.20 \div 0.14$ &$-0.015 \div 0.090$ &$-0.20 \div 0.29$
&$-0.028 \div 0.033$ & $-0.05 \div 0.05$ & 0 \\
\cline{2-6}
   &$-0.067 \div 0.047$&$-0.003 \div 0.025$&$-0.06 \div 0.04$ &
 $-0.006 \div 0.007$ & $-0.01 \div 0.01$ & \\
\hline\hline
% 41
  & $-0.47~ \div 0.40~$ & $-0.42~ \div 1.05~$ &
& $-0.54~ \div 0.74~$ & $ -0.88~ \div 0.88$ &  \\
\cline{2-3} \cline{5-6}
 4 & $-0.046 \div 0.036$ &$-0.013 \div 0.083$
&
$\displaystyle - {s_W\over c_W} x_\gamma$ & $-0.027 \div 0.031$
& $-0.05 \div 0.05$
& 0 \\
\cline{2-3} \cline{5-6}
   &$-0.012 \div 0.014$ &$-0.003 \div 0.003$ & &$-0.006 \div 0.006$
&$-0.01 \div 0.01$ & \\
\hline\hline
% 32
  & $-0.45~ \div 0.31~$ & $-0.31~ \div 0.96~$ & & $-0.20~ \div 0.24~$ & & \\
\cline{2-3} \cline{5-5}
 3 & $-0.044 \div 0.028$ & $-0.011 \div 0.080$ &
$\displaystyle -{s_W\over c_W} x_\gamma$
& $-0.010 \div 0.011$ &
$\displaystyle {c_W\over s_W} y_\gamma$ & 0 \\
\cline{2-3} \cline{5-5}
   & $-0.011 \div 0.011$ &$-0.003 \div 0.022$ & &$-0.002 \div 0.002$ & & \\
\hline\hline
% 22
  & & $-0.09~ \div 0.10~$ &  & $-0.17~ \div 0.19~$ & & \\
\cline{3-3} \cline{5-5}
 2 &
$\displaystyle {x_\gamma\over {s_W c_W}}$ &$-0.004 \div 0.005$
&
$\displaystyle -{s_W \over c_W} x_\gamma$
&$ -0.009 \div 0.009$ &
$\displaystyle {c_W\over s_W} y_\gamma$  &  0 \\
\cline{3-3} \cline{5-5}
   & &$ -0.001 \div 0.001$ & &$-0.002 \div 0.002$ & & \\
\hline\hline
% 16
  & & & & $-0.07~ \div 0.08~$ & & \\
\cline{5-5}
 1 & 0 & 0 & 0 & $-0.006 \div 0.008$ &
$\displaystyle {c_W \over s_W} y_\gamma$ & 0 \\
\cline{5-5}
   & & & & $-0.002 \div 0.002$ & & \\
\hline\hline\hline
% 61
  &$-5.09~ \div 4.82~$ & $-0.47~ \div 1.14~$ & $-7.28 \div 7.53$ &
$-0.58~ \div 0.82~$ & $-2.23 \div 2.33$ & $\pm 0.2~~~$ \\
\cline{2-7}
 6 &$-0.22~ \div 0.21~$ &$-0.015 \div 0.093$ &$-0.26 \div 0.31$
 &$-0.029 \div 0.034$ & $-0.05 \div 0.05$ & $\pm 0.005~$ \\
\cline{2-7}
   &$-0.075 \div 0.083$ &$-0.003 \div 0.026$ &$-0.07 \div 0.07$
&$-0.007 \div 0.007$ &$-0.01 \div 0.01$ &$ \pm 0.0005$ \\
\hline
\end{tabular}
\caption[{\bf Table 5}]
{\it The $95 \% C.L.$ bounds on the non-standard trilinear couplings
obtained from fits.
For each set of free parameters, we present
the results of the fits carried out
at $\sqrt{s}=190GeV,~500GeV$ and $1000GeV$ for the respective
luminosities in Table 4. Note that the various choices of free parameters
and constraints correspond to the cases listed in Table 1.}
{\label{tab5}}
\end{center}
\end{table}
%-----------------------------------------------------------------------------
\normalsize
and
contour plots are presented for two different
two-parameter fits in figs. 1a,b
and for a three-parameter fit in figs. 2a,b,c.
\par
According to Table 5,
the absolute values of the bounds (for one-parameter cases)
at $500GeV$ reach the order
of magnitude of the standard
radiative corrections, which, when represented in
terms of the couplings $\dz,~\xg,~\xz$, etc. \cite{RC92}, are
of the order of 0.01 to 0.001. For the beam energy $1000GeV$ such an
accuracy can even be reached in certain multi-parameter cases.
The strong bound on the anapole coupling,
$\zz$, even in the presence of all other
non-standard terms, is related to the very
particular helicity dependence
of the anapole interaction (see Table 2).
\par
The explicit numerical results in Table 5 are in very good agreement with
the predictions of the scaling law given by eqs. \rfn{3.4},\rfn{3.5}
and explicitly evaluated in Tables 3 and 4. Even though the scaling
laws are based on the high-energy limit, $s >> 4 M^2_W$, the
numerical analysis shows their validity even when the LEP2 results
at $190 GeV$ are used as starting point.
\par
%\footnotesize
The disappearance of $(\xg,~\yg)$-correlations with
increasing energy in fig. 1b is related to the fact that the
contribution to the transverse-transverse cross section
of $\dz=\xg /s_Wc_W$ (according to Table 2)
decreases asymptotically as $1/s^2$ and becomes
negligible, thus allowing for a clear separation
of $\xg$ and $\yg$, as in this limit these parameters
contribute to different
helicity amplitudes only. In contrast,
as seen from fig. 1a,
the correlations
between the dimension-4 couplings, $\dz,\xg,\xz$, become stronger
at high energies.

%\newpage
\mysection{Conclusions}
In the present work, we have extended our LEP 2 analysis on
$e^+e^- \to W^+W^-$
to the energy range of a future $e^+e^-$ linear collider working at
$500 GeV$
to $1000 GeV$. For integrated luminosities of
$10 fb^{-1}$ and $44 fb^{-1}$
at $500 GeV$ and $1000 GeV$, respectively, we found that the bounds
on non-standard couplings will be of the order
of $10^{-2}$ to
$10^{-3}$.
With respect to measurements at LEP2, this will be an improvement
by at least one order of magnitude. In fact, the bounds will reach
the magnitude
of (standard) radiative corrections.
\par
We have derived simple scaling laws for the sensitivity of the reaction
$e^+e^- \to W^+W^-$ for the determination of
non-standard couplings. According to the scaling laws the sensitivity
increases as $s\sqrt L$ for the anapole and as $\sqrt{sL}$
for all other couplings.
This agrees with what we have found in
the detailed simulation of the analysis of future data.
\vskip 3cm\noindent
{\bf Acknowledgement}
\vskip 0.5cm\noindent
We would like to thank S. Katsanevas for the useful discussions.
\normalsize

\vfil\eject

\section*{Figure captions}

{\bf Figure 1:}
\hspace{0.5cm}
Contour plots (95\% C.L.)
obtained in the two-parameter fits of\\
\hspace*{3cm}
a) the parameters $\dz$ and $\xg$ with the constraints
$\xz = - {s_W \over c_W} \xg~~,~~\yg = \yz = 0,$\\
\hspace*{3cm}
b) the parameters $\xg$ and $\yg$ with the constraints\\
\hspace*{4cm}
$\dz={\xg \over {s_Wc_W}}~~,~~\xz = - { s_W \over c_W} \xg~~,~~
\yz = {c_W \over s_W} \yg$.\\
\par\noindent
{\bf Figure 2:}
\hspace{0.5cm}
Contour plots (95\%~C.L.) for the three-free-parameter case,
($\dz,~\xg,~\yg$),\\
\hspace*{3cm}
with the constraints
$\xz=-{s_W \over c_W} \xg~~,~~\yz = {c_W \over s_W} \yg$~~ in\\
\hspace*{3cm}
a) the
$ (\xg,~\dz)$ plane,\\
\hspace*{3cm}
b) the
$(\yg,~\dz)$ plane,\\
\hspace*{3cm}
c) the
$(\xg,~\yg)$ plane.

\end{document}